\documentclass[12pt]{article}
\usepackage{graphics}
\usepackage{graphicx}

\input{epsf}
\usepackage{cite}
\def\@fmsl@sh#1#2#3{\m@th\ooalign{$\hfil#1\mkern#2/\hfil$\crcr$#1#3$}}
 \def\eq#1\en{\begin{equation}#1\end{equation}}
\def\s[#1,#2]{[#1\stackrel{\star}{,}#2]}
\def\sx[#1,#2]{[#1\stackrel{\star_{x}}{,}#2]}

\newcommand{\nc}{\newcommand}
\nc{\beq}{\begin{equation}}
\nc{\eeq}{\end{equation}}
\nc{\beqa}{\begin{eqnarray}}
\nc{\eeqa}{\end{eqnarray}}

\def\gsim{\mathrel{\rlap{\lower4pt\hbox{\hskip1pt$\sim$}}
    \raise1pt\hbox{$>$}}}       

\begin{document}
\makeatletter
\def\fmslash{\@ifnextchar[{\fmsl@sh}{\fmsl@sh[0mu]}}
\def\fmsl@sh[#1]#2{%
  \mathchoice
    {\@fmsl@sh\displaystyle{#1}{#2}}%
    {\@fmsl@sh\textstyle{#1}{#2}}%
    {\@fmsl@sh\scriptstyle{#1}{#2}}%
    {\@fmsl@sh\scriptscriptstyle{#1}{#2}}}
\def\@fmsl@sh#1#2#3{\m@th\ooalign{$\hfil#1\mkern#2/\hfil$\crcr$#1#3$}}
\makeatother


\title{\large{\bf Invisible Higgs boson, continuous mass fields and unHiggs mechanism}}

\author{X. ~Calmet$^a$\thanks{Charg\'e de recherches du F.R.S.-FNRS.}
\thanks{xavier.calmet@uclouvain.be}, N.~ G.~Deshpande$^b$\thanks{desh@uoregon.edu},
X.~G.~He$^c$\thanks{hexg@phys.ntu.edu.tw},
S.~D.~H.~Hsu$^b$\thanks{hsu@uoregon.edu}\\
$^a$Catholic University of Louvain,
Center for Particle Physics and Phenomenology,\\
2, Chemin du Cyclotron,
B-1348 Louvain-la-Neuve, Belgium
\\
$^b$Institute of Theoretical Science, University of Oregon,\\
Eugene, OR 97403 USA\\
$^c$Department of Physics and Center for Theoretical Sciences,\\
National Taiwan University, Taipei, Taiwan, R.O.C.
}
\date{October, 2008}

\maketitle

\begin{abstract}
We explore the consequences of an electroweak symmetry breaking sector which exhibits approximately scale invariant
dynamics -- i.e., nontrivial fixed point behavior, as in unparticle
models. One can think of an unHiggs as a composite Higgs boson with a continuous mass distribution.  We find it convenient to represent the unHiggs in terms of a
K\" allen-Lehmann spectral function, from which it is simple to verify
the generation of gauge boson and fermion masses, and unitarization
of WW scattering. We show that a spectral function with broad
support, which corresponds to approximate fixed point behavior over an
extended range of energy, can lead to an effectively invisible Higgs
particle, whose decays at LEP or LHC could be obscured by background.  
\end{abstract}


\newpage

\bigskip
Recently there has been significant interest in the possibility of an unparticle sector of fundamental physics which is approximately scale invariant \cite{unparticle}. Most models have assumed that the unparticle sector is peripheral to the standard model, but recently Stancato and Terning \cite{Stancato:2008mp} have considered the possibility that the sector that spontaneously breaks electroweak symmetry is approximately scale invariant, leading to an unHiggs boson,  see ref. \cite{grave} for works on related ideas. In \cite{Deshpande:2008ra} it was shown that scale invariance can be described in terms of particles with continuous masses \cite{conti} or, equivalently, with more complicated than usual K\" allen-Lehman representation \cite{KL}. In this letter we apply the continuous mass formalism to the unHiggs, deducing rather simply how fermion masses are generated, how unitarity is preserved in the presence of massive gauge bosons, and the form of radiative corrections. Further, we illustrate that if scale invariance holds over a large range of energies the resulting unHiggs particle is effectively a broad resonance, which may be extremely difficult to detect. In this scenario of a potentially effectively invisible unHiggs there is no violation of unitarity and no disagreement with electroweak precision data, yet no Higgs would be seen at LHC.

To illustrate the basic mechanism we consider a scalar field with a continuous mass $\phi(x,\rho)$. The corresponding unparticle field $\phi_{\cal U}$ is defined as in \cite{Deshpande:2008ra}:
\begin{eqnarray} \label{UH}
\phi_{\cal U}(x)= \int_0^\infty \phi(x,\rho) f(\rho) d\rho
\end{eqnarray}
where $f(\rho)=a_d \rho^{d/2-1}$ with 
\begin{eqnarray}
a_d^2= \frac{{A}_d}{2 \pi}, \ \ {A}_d = \frac{16 \pi^{5/2} \Gamma(d+1/2)}{(2\pi)^{2d} \Gamma(d-1) \Gamma(2d)}.
\end{eqnarray}
By choosing the continuous mass field with appropriate gauge properties we can use it to implement symmetry breaking. The field $\phi(x,\rho)$ is chosen to be dimensionless. As an example, we begin by assuming that $\phi$ is charged under a U(1) gauge symmetry.  One could trivially generalize our consideration to any non-abelian gauge group. We consider the following Lagrangian density which has a U(1) gauge invariance in the $x$-space:
\begin{eqnarray}
\label{L}
{\cal L}(x)&=& \int^\infty_0 \left(  D_\mu \phi(x,\rho)^*  D^\mu \phi(x,\rho) +\rho \phi^*(x,\rho)\phi(x,\rho)-\lambda(\rho) 
(\phi^*(x,\rho)\phi(x,\rho))^2 \right) ~d\rho  \\ && \nonumber - \frac{1}{4} F_{\mu\nu}(x)F^{\mu\nu}(x),
\end{eqnarray}
where $\lambda$ and $\rho$ have mass dimension +2 and the scalar $\phi$ is dimensionless. The covariant derivative is given by $D_\mu=\partial_\mu+i g A_\mu(x)$, note that $A_\mu$ is only a function of $x$ and not $\rho$.
Under  local U(1) gauge transformations one has, as usual,
\begin{eqnarray} 
\phi^\prime(x,\rho)= e^{i \alpha(x)} ~\phi(x,\rho) \\
A^\prime_\mu(x)= A_\mu(x)- \frac{1}{g}\partial_\mu \alpha(x).
\end{eqnarray}
In the limit where $\lambda = 0$ the action is scale invariant: under a scale transformation $x \rightarrow \Lambda^{-1} x$, $\rho \rightarrow \Lambda^2 \rho$, the Lagrangian density is rescaled by $\Lambda^4$, so that $S = \int d^4x~ {\cal L}(x)$ is invariant. Note the importance of the limits of integration $0 \leq \rho \leq \infty$ in this result. If instead the range of integration is finite, scale invariance is broken. Similarly, the interaction $\lambda \phi^4$ in general breaks scale invariance, unless $\lambda$ is proportional to $\rho$.

Note that in this formalism the path integral quantization of the field $\phi (x, \rho)$ requires the measure $\prod_{x \rho} d\phi (x, \rho)$, so from this perspective there are an infinite number of new degrees of freedom. Similarly, the canonical quantization conditions are imposed on $\phi (x, \rho)$ for each value of $\rho$. In a microphysical realization, e.g., in a confining strongly coupled gauge model, the scalar unparticle corresponds to a particular convolution of $\phi(x,\rho)$, and the continuous mass formalism is simply a model for the behavior of the unparticle; in particular, it reproduces the correct propagator and scaling dimension. In that context the additional degrees of freedom, beyond the special convolution, are not regarded as physical degrees of freedom. The unparticle bound state arises from a finite number of short distance degrees of freedom, whose dynamics fix the values of the functions $\lambda (\rho)$, etc. The confining theory could be a Banks-Zaks model \cite{Banks:1981nn} in which case the fixed point behavior, which presumably holds over some range in energy, fixes the limits of the integral over $\rho$ to some range $\rho_1 \leq \rho \leq \rho_2$. Presumably, $\rho_2 \gg \rho_1$ so that the scale invariance that applies when the limits are zero and infinity is approximately true for momenta in the fixed point region. If $\rho_1 \rightarrow 0$ very strict limits on unparticles arise due to the long range forces they mediate \cite{lrf}. 
Clearly there are challenges in assuming the existence of a confining gauge theory sector, some of whose matter degrees of freedom carry SU(2)$_L$ and condense to form the Higgs. We leave aside those model building issues and concentrate on the phenomenology of an unHiggs. For examples of dynamical models which might realize a light composite Higgs, see, e.g., \cite{compositehiggs}.

The vacuum expectation value of the field $\phi(x,\rho)$ is given by
\begin{eqnarray}
\label{vev}
v(\rho)=\sqrt{\frac{\rho}{2\lambda(\rho)}},
\end{eqnarray}
and we denote the fluctuation around $v(\rho)$ by $h(x,\rho)$. The mass of the gauge boson after spontaneous symmetry breaking can be seen from Eq.~(\ref{L}) to be 
\begin{eqnarray}
m_A^2 =\frac{1}{4} g^2 \int d\rho ~v(\rho)^2 
\end{eqnarray}
and is independent on $\rho$. Presumably, we would like to set the lower limit of $\rho$ integration to be larger than the $Z$ mass (or the weak scale), in order to have a low energy effective theory with a scalar degree of freedom which is a bound state.  The mass $m(\rho)$ of the field $h(x,\rho)$ is given by
\begin{eqnarray}
m^2 (\rho)= 2 \rho.
\end{eqnarray}

If we extend our U(1) continuous mass Higgs model to non-abelian groups and in particular to the standard model, the couplings of the Higgs to the gauge bosons is modified, the two gauge bosons Higgs coupling is given by
\begin{eqnarray} \label{coupling}
g^2 A_\mu A^\mu \int d\rho ~v(\rho) h(x,\rho),
\end{eqnarray}
where $h(x,\rho)$ is the fluctuation around the vacuum expectation value and we have suppressed the group indices. If we were to take the continuous mass theory literally, only one particular convolution of the field is eaten, leaving an infinite number of additional degrees of freedom that couple to the gauge bosons. If the continuous mass theory is used only as a model for an unHiggs bound state, those additional degrees of freedom are fictitious. In particular, only three Goldstone modes result from the physical convolution, and those are eaten by the $W^{\pm}$ and Z in the standard model.

The Yukawa couplings are of the form
\begin{eqnarray}
\int d\rho~Y(\rho) \bar \Psi_L(x) H(x,\rho)  \Psi_R(x) ~+~ h.c.
\end{eqnarray}
where $H(x,\rho)$ is the Higgs doublet and $Y(\rho)$ has mass dimension -1. Note that the Yukawa couplings are not necessarily $\rho$ dependent. One can write  $Y(\rho)=\tilde Y / \sqrt{\rho}$ and rescale $H(x,\rho)$ to obtain a $\rho$ independent Yukawa coupling.  In general, unless a specific form is assumed for the Yukawa constant $Y(\rho)$, Yukawa couplings break conformal invariance.

The propagator for the field $h(x,\rho)$ has been evaluated in \cite{Deshpande:2008ra} and is given by
\begin{eqnarray}
\int d^4x ~e^{i p x}  \langle 0|T h(x, \rho) h(0, \rho^\prime)|0 \rangle ~=~ \frac{i}{p^2-m^2(\rho)+i\epsilon} \delta(\rho-\rho^\prime).
\end{eqnarray}
Note that this is essentially a K\" allen-Lehmann propagator \cite{KL}:
\begin{eqnarray}
\Delta_{\rho \rho'} (p)= \int_0^\infty \frac{i}{p^2-\mu^2+i\epsilon} \Omega_{\rho \rho'} (\mu^2) d\mu^2,
\end{eqnarray}
with a spectral function $\Omega_{\rho \rho'} (\mu^2) = \delta (\mu^2 - m^2(\rho)) \, \delta (\rho - \rho')$. 

In our formalism the unHiggs coupling to two gauge bosons is given by Eq.~(\ref{coupling}), which yields an unHiggs boson
\begin{equation}
\phi_{\cal U} (x) = \int d\rho ~v (\rho) h (x, \rho)
\end{equation}
with propagator
\begin{equation}
\label{uprop}
\Delta_{\cal U} (p) = \int d\rho~ \frac{v^2 (\rho)}{p^2 - m^2 (\rho) + i \epsilon}.
\end{equation}
The scaling properties of $\phi_U$ depend on the scaling properties of $v (\rho)$, which in turn depend on $\lambda (\rho)$. The choice $\lambda (\rho) = c \rho$ preserves scale invariance, leading to constant $v (\rho)$ and unHiggs scaling dimension $d=2$. In general, however, $\lambda (\rho)$ can have any functional form and we can have unHiggs of arbitrary dimension. Fermion masses and Yukawa couplings are given by
\begin{equation}
m_f = \int d\rho ~Y (\rho) v (\rho)
\end{equation}
and
\begin{equation}
\int d\rho ~Y(\rho) \bar \Psi_L(x) h(x,\rho)  \Psi_R(x) +h.c.~
\end{equation}
In order to preserve the property that only one particular convolution of the continuous mass field is physical, we must choose the Yukawa coupling function $Y(\rho)$ proportional to $v(\rho)$ such that the same convolution couples to fermions and gauge bosons. The constant of proportionality is $g^2 m_f  / 4 m_W^2$, and thus uniquely defined for each fermion.

The Higgs mechanism for a continuous mass field does not lead to a violation of unitarity of the S-matrix if most of the mass of the Higgs is concentrated below 1 TeV. Since the gauge symmetry is spontaneously broken by a Higgs mechanism, which is an low-energy effect of the vacuum state, we expect the high energy behavior of the model should still be that of an unbroken gauge theory. Indeed it is easy to show using the result of \cite{Duncan:1985vj} that  the contribution of the unHiggs to WW elastic scattering is given by:
\begin{eqnarray}
A_{sH} =\frac{-i g^4 s^2}{64m_W^4} (1+\beta^2)^2 \int d\rho~ \frac{v(\rho)^2}{s-m^2(\rho)}
\end{eqnarray}
\begin{eqnarray}
A_{tH} =\frac{-i g^4 s^2}{64m_W^4} (\beta^2-\cos\theta)^2 \int d\rho~ \frac{v(\rho)^2}{t-m^2(\rho)}
\end{eqnarray}
where $\beta=(1-4/s)^{1/2}$ and $\theta$ is the scattering angle. In these expressions the usual Higgs propagator is replaced by the unHiggs propagator. Note that in the limit $s,t \gg \rho$, we recover the standard model result. As long as the range of integration terminates at a value not much greater than the 1 TeV unitarity bound \cite{Lee:1977eg}, the unHiggs boson unitarizes the amplitude of the elastic WW scattering. Note that in the approach of \cite{Stancato:2008mp} it is nontrivial to verify unitarization.

We shall now calculate the production cross-section of the unHiggs in a lepton collider such as LEP. The dominant mode at LEP for the production of a light Higgs was via Higgs--strahlung. The production cross-section via unHiggs--strahlung at an $e^+e^-$ machine is given by
\begin{eqnarray}
\sigma(e^+e^- \to HZ)&=& \frac{g^2}{4 m^2_W}\frac{\pi\alpha^2}{24} \frac{(1-4 x_W+8 x_W^2)}{x_W^2(1-x_W)^2} \\ \nonumber && 
\times  \int_{\rho_1}^{\rho_2} d\rho ~v(\rho)^2 ~\frac{2K(\rho)}{\sqrt{s}} 
\frac{(K(\rho)^2+3m_Z^2)}{(s-m_Z^2)^2}
\end{eqnarray}
where $x_W=\sin^2\theta_W$ and 
\begin{eqnarray}
K(\rho)=\frac{\sqrt{s}}{2} ~ \sqrt{1-\frac{2}{s}(m_h^2(\rho)+m_Z^2) + \frac{(m_h^2(\rho)-m_Z^2)^2}{s^2}} ~~,
\end{eqnarray}
where $m_h^2 (\rho) = 2\rho$. If the $Z$ boson is off-shell, $m_Z$ in the $K(\rho)$ is replaced by the four momentum squared of the $Z$-boson. The unHiggs could behave as a very broad Higgs boson since its mass could be distributed over a large energy spectrum. Note that the production cross-section into each energy bin could be much smaller than in the case where the standard model Higgs has that particular mass. This is similar to the results of van der Bij et al. \cite{vdB}, who first identified a number of ways that LEP could have missed the Higgs boson. If the mass is spread between, for example, 90 GeV and 115 GeV, the unHiggs could easily have escaped detection at LEP. Similarly a sufficiently broad, perhaps heavier, Higgs would be difficult to observe at the LHC.

Finally, we calculate the contribution to the $S$ parameter from the unHiggs relative to that of a reference standard model Higgs. It is given by
\begin{eqnarray}
S ~\approx~ \frac{g^2}{4m^2_W}
 \, \int d\rho ~ \frac{1}{12 \pi} v^2(\rho)  \log\left(\frac{m_h^2(\rho)}{m_{H,ref}^2}
\right),
\end{eqnarray}
where we assume that $m_h(\rho) \gg m_W$ as in \cite{Peskin:1991sw}, $m_{H,ref}$ is a standard model Higgs reference mass, and our $S$ is defined relative to that value. If we take $\rho_1$ and $\rho_2$ less than 115 GeV we can obtain a  better fit than the standard model one with a Higgs mass greater than 115 GeV, although there are probably model building challenges to extending scale invariance down to such low energies. 

Note our results are valid for unHiggses of arbitrary scaling dimension. If we choose
\begin{equation}
\lambda (\rho) = \frac{\rho}{C^2} \left( \frac{\Lambda^{d-2}}{f(\rho)^2} \right)~~,
\end{equation}
where $C$ is a dimensionless constant and $f (\rho)$ is defined below Eq.~(\ref{UH}), then the unHiggs coupling to gauge bosons is given, using (\ref{vev}), by
\begin{equation}
\label{uAA}
\sim g^2 A_\mu A^\mu \frac{C}{\Lambda^{\frac{d-2}{2}}} \int d\rho~ f(\rho) h (x ,\rho)~~,
\end{equation}
which describes an unHiggs of dimension $d$. The consequences of such a choice are obtained simply by replacing $v (\rho)$ by $C f(\rho) / \Lambda^{\frac{d-2}{2}}$. The value of $C$ should be of order unity and the scale $\Lambda$ a few hundred GeV.

\bigskip

In this note we have explored the phenomenology of an unHiggs mechanism, in which electroweak symmetry is broken by a field with approximate scale invariance. Using our continuous mass formalism, it is easy to deduce many of the properties of an unHiggs. In essence, the unHiggs would behave as a very broad resonance with the usual Higgs interactions. However, because any signals it produces are spread over a large range in energy the unHiggs can be hidden by background processes. 

Our formulation is quite different from that in \cite{Stancato:2008mp}. The central object in our analysis is the continuous mass field $\phi(x,\rho)$, which has the SU(2)$\times$U(1) quantum numbers of the usual Higgs. We implement spontaneous symmetry breaking by causing $\phi(x,\rho)$ to obtain a vacuum expectation value. In this approach unitarization is automatic, since we have clearly only spontaneously broken the gauge symmetry; the high energy behavior of the model should be unaffected. The specific unparticle properties, such as the scaling dimension $d$, are obtained by choosing the appropriate function $\lambda (\rho)$, which determines $v (\rho)$, and leads to the desired propagator as in Eq.~(\ref{uprop}), and the appropriate coupling to gauge bosons as in Eq.~(\ref{uAA}).

We have not discussed the underlying dynamical model for this mechanism, but it would presumably require strong dynamics, a fixed point, perhaps of the Banks-Zaks type, and additional particles, some of which must carry SU(2)$_L$ and hypercharge quantum numbers.

\bigskip

\emph{Acknowledgments---}   The work of X.C. is supported in part by the Belgian Federal Office for Scientific, Technical and Cultural Affairs through the Interuniversity Attraction Pole P6/11. N.D. and S.H. are supported by the Department of Energy under DE-FG02-96ER40969. X.G.H. is supported by NSC and NCTS.

\bigskip
\bigskip



\bigskip


\end{document}